# Intentionally Unintentional: GenAI Exceptionalism and the First Amendment

David Atkinson,[1] Jena D. Hwang,[2] and Jacob Morrison[3]

## ABSTRACT

This paper challenges the assumption that courts should grant First Amendment protections to outputs from large generative AI models, such as GPT-4 and Gemini. We argue that because these models lack intentionality, their outputs do not constitute speech as understood in the context of established legal precedent, so there can be no speech to protect. Furthermore, if the model outputs are not speech, users cannot claim a First Amendment speech right to receive the outputs. We also argue that extending First Amendment rights to AI models would not serve the fundamental purposes of free speech, such as promoting a marketplace of ideas, facilitating self-governance, or fostering self-expression. In fact, granting First Amendment protections to AI models would be detrimental to society because it would hinder the government's ability to regulate these powerful technologies effectively, potentially leading to the unchecked spread of misinformation and other harms.

## I. INTRODUCTION

Since ChatGPT first burst into society's collective consciousness toward the end of 2022, scholars have pondered its implications. While discussions of copyright receive the lion's share of attention,[4] and privacy rights consume most of the remaining spotlight,[5] the debate of whether generative AI (GenAI) models should

---

[4] Katherine Lee, A. Feder Cooper, & James Grimmelmann, *Talkin' 'Bout AI Generation: Copyright and the Generative-AI Supply Chain*, CS&LAW '24: PROC. OF THE SYMP. ON COMPUT. SCI. AND L. (2024).
[5] *See, e.g.*, Chen Ruizhe et al., Learnable Privacy Neurons Localization in Language Models (May 16, 2024) (unpublished manuscript) (https://arxiv.org/abs/2405.10989); Zhipeng Wang et al., Information Leakage from Embedding in Large Language Models (May 22, 2024) (unpublished manuscript) (https://arxiv.org/abs/2405.11916).

receive First Amendment protections is growing.[6]

It bears emphasizing that whether First Amendment protections apply to GenAI outputs is an unsettled legal question. The importance of this analysis can't be overstated given the convergence of several factors: (1) the number of GenAI models like GPT-4 and Gemini is proliferating, (2) the use of GenAI in everyday life is becoming more common, and (3) people tend to anthropomorphize things that seem to have human characteristics.[7]

Additionally, the legal implications of assigning First Amendment protection to GenAI, in effect, would necessarily deem its outputs as speech. The repercussions of doing so would be non-trivial, as legal scholars Karl Manheim and Jeffrey Atik explain:

> [If GenAI output is speech, it] would likely prohibit treating AI as a product or attaching liability to harmful outputs. It could also give AI companies free rein to collect and use personal information as data inputs for their algorithmic (constitutionally protected) outputs.[8] It is not just privacy rights that would vanish under such a regime, but many forms of consumer protection and other regulatory objectives.[9]

The potential consequences of assigning protections to GenAI outputs rest on the premise that GenAI models are constitutionally recognizable speakers, and their outputs are, therefore, speech. Works criticizing this premise have pointed out GenAI's tendencies to produce nonsense or asemic language.[10] Others have contended that GenAI cannot be a speaker as it does not participate in communication.[11]

---

[6] *See, e.g.*, Cass R. Sunstein, Artificial Intelligence and the First Amendment (April 28, 2023) (unpublished manuscript) (https://papers.ssrn.com/sol3/papers.cfm?abstract_id=4431251); Eugene Volokh, Mark A. Lemley, & Peter Henderson, *Freedom of Speech and AI Output*, 3 J. FREE SPEECH L. 651 (2023).

[7] The way we talk about chatbots warps our understanding. We "chat" with it and have "conversations." It "responds" based on what it "knows." Since we, as humans evolved over millennia to socialize with others, tend to interpret the outputs of the model in that social context, we also naturally extend the metaphor to believing the model is providing meaningful outputs as expected from another human being.

[8] *See* NetChoice v. Bonta, 692 F.Supp.3d 924, 936–37 (N.D. Cal. 2023) (issuing a preliminary injunction against the California Age-Appropriate Design Code Act on the ground that regulating the collection and use of children's personal information infringed the free speech rights of online tech companies).

[9] Karl Manheim & Jeffery Atik, *AI Outputs and the Limited Reach of the First Amendment*, 63 WASHBURN L. J. 159, 161 (2023).

We draw from and build on these works to argue that there are no constitutionally recognized speakers in GenAI because, unlike human beings, models lack communicative intent, akin to stochastic parrots.[12] This, in turn, means there is no speech that the First Amendment can protect. Moreover, if there is no speech, there can be no constitutionally recognized listeners of speech, meaning users of the models generally have either weak or no First Amendment right to receive any model outputs. Consequently, if the First Amendment does not apply, the government can more freely regulate GenAI. We further argue that GenAI outputs are not speech; therefore, no court should take the extraordinary step of extending the highest First Amendment protections and scrutiny to non-speech by non-humans.

Our primary contributions to this discussion include clarifying the legal stakes (such as the unprecedented expansion of free speech rights to a non-human entity) and the most common arguments involved in First Amendment discussions regarding GenAI outputs (like whether the developer, model, or recipients have speech rights that would likely trigger strict scrutiny) by unifying multiple concepts within legal theories, including asemic language, intentionality, human involvement, generation versus use, stochasticity, and substantive constitutional arguments. This research also provides an informed technical description by including AI researchers as co-authors. Finally, the brevity of this paper compared to comparable papers facilitates a concise discussion using accessible prose to make concepts more broadly understandable beyond legal scholars.

*A. Scope*

---

[10] *E.g.*, Dan L. Burk, *Asemic Defamation, or, the Death of the AI Speaker*, 22 FIRST AMEND. L. REV. 189 (2024).

[11] *E.g.*, Manheim & Atik, *supra* note 9.

[12] Emily M. Bender et al., *On the Dangers of Stochastic Parrots: Can Language Models Be Too Big?*, PROC. OF THE 2021 ACM CONF. ON FAIRNESS, ACCOUNTABILITY, AND TRANSPARENCY 616 (2021).

There are some important limitations of the following argument: (1) The argument only considers foundation models like GPT-4, Claude, Gemini, and Llama, not all GenAI models, and (2) relatedly, the argument asserts that not all model outputs are protected by the First Amendment, not that no outputs could ever be protected by the First Amendment.[13] Generally, the more directed the model is, the more likely that some protections may attach.[14] Our default argument for all foundation models is that virtually none of their outputs are protected speech.

*B. How GenAI Works*

Before going any further, it is helpful to understand how GenAI functions. In a few words, GenAI models are mathematical functions for predicting what words follow a given input. Models do this by inputting tokens (subcomponents of words) from the prompt into a model, taking its output as the token most likely to follow the input, and repeating this process until the desired output length is achieved.

Tokens are created by a "tokenizer," which is a model that takes in data and separates it into tokens, which are then fed into a model. A token is a series of weights that define the subcomponent of a word, and the weights within the token help define the statistical relationship between different tokens learned during training.[15] On the other hand, the weights within a model are tuned during training to take in tokens and "learn" relationships between them to predict likely outputs. Its primary function, in other words, is to respond to user prompts with plausible-sounding outputs based on what the model was trained to "understand" as likely related tokens. The models can generalize to some extent,[16] and even without referring

---

[13] It seems at least theoretically possible that one day AI could become self-aware, and if so, it could make expressions with intention. But that is not the case today, and there is no reason to believe it will be the case in the near future. Regardless, the First Amendment only applies to humans, another requirement for protected speech.

[14] Outputs may only be speech when the developer directs the model to produce a particular or distinct output. Merely applying an insignificant filter does not automatically turn non-speech into speech, as that would make it trivially easy to make all outputs protected speech, subverting the purpose of the First Amendment.

[15] Similar to John Rupert Firth's "You shall know a word by the company it keeps." J.R. Firth, *A Synopsis of Linguistic Theory, 1930–1955, in* STUDIES IN LINGUISTIC ANALYSIS 11 (Oxford 1962).

[16] Some argue it's not true generalization but is instead something like "approximate retrieval". *See, e.g.*,

to an external database they can sometimes memorize and output images and text that is a verbatim or near-verbatim version of the material it was trained on, perhaps indicating the models themselves are a kind of database.

## C. The First Amendment

And for the last introductory matter, it is helpful to understand the nature of speech protected by the First Amendment. The Amendment reads:

> Congress shall make no law respecting an establishment of religion, or prohibiting the free exercise thereof; or abridging the freedom of speech, or of the press; or the right of the people peaceably to assemble, and to petition the Government for a redress of grievances.[17]

We bring attention to the following portion that focuses on speech: "Congress shall make no law . . . abridging the freedom of speech, or of the press[.]"

The Supreme Court has interpreted the First Amendment's freedom of speech clause in a long series of cases. This paper will explore a few relevant quotes later, but for now, it is sufficient to think of it as protecting the "marketplace of ideas," self-governance, and self-expression. The protections extend not only to speakers (the people producing the speech) but also to listeners (the people receiving the speech), because it would undermine the purposes of the First Amendment to allow anyone to say anything but bar others from receiving the communication.

Finally—and this is worth stating clearly—there is no binding case law in the United States that grants First Amendment speech protections to anything that was created absent a human's significant, intentional

---

Zhaofeng Wu et al., *Reasoning or Reciting? Exploring the Capabilities and Limitations of Language Models Through Counterfactual Tasks*, CONF. OF THE N. AM. CHAPTER OF THE ASS'N FOR COMP. LINGUISTICS 1819–62 (2024).

[17] U.S. CONST. amend. I.

involvement.[18]

## II. THE CODE IS SPEECH, SO MODELS ARE SPEECH ARGUMENT[19]

The first argument many will make is that GenAI functions like code, and code is protected speech; therefore, GenAI outputs are protected speech. This notion is misguided, but it is worth walking through the argument to better understand why the argument is misplaced.

There are at least three cases in multiple circuits where the courts determined that code is speech protected by the First Amendment. The courts decided each case around the turn of the century as software became an unavoidable legal topic with the widespread use of the internet. In two of the three cases (*Junger v. Daley*[20] and *Bernstein v. United States Department of Justice*[21]), the issue was government-export control restrictions that attempted to prevent researchers from sharing cryptographic code.

For example, in *Junger v. Daley*, the Sixth Circuit found that "Because computer source code is an expressive means for the exchange of information and ideas about computer programming, we hold that it is protected by the First Amendment."[22]

Likewise, in *Bernstein*, the Ninth Circuit held that:

[C]ryptographers use source code to express their scientific ideas in much the same way that

---

[18] *See, e.g.*, Miles v. City Council, 710 F.2d 1542, 1544 n.5 (11th Cir. 1983) ("This Court will not hear a claim that Blackie's [the cat's] right to free speech has been infringed. First, although Blackie arguably possesses a very unusual ability, *he cannot be considered a "person" and is therefore not protected by the Bill of Rights*. Second, even if Blackie had such a right, we see no need for appellants to assert his right *jus tertii*. Blackie can clearly speak for himself." (emphasis added)).

[19] The structure of this section of the paper was informed by comments from the Center for Democracy & Technology. Center for Democracy & Technology, *Re: NTIA's Request for Comment Regarding Dual-Use Foundation Artificial Intelligence Models with Widely Available Model Weights as per Section 4.6 of the Executive Order on the Safe, Secure, and Trustworthy Development and Use of Artificial Intelligence*, CDT.ORG (Mar. 27, 2024), https://cdt.org/wp-content/uploads/2024/03/CDT-to-NTIA-comments-on-open-foundation-models-03272023.pdf.

[20] Junger v. Daley 209 F.3d 481 (6th Cir. 2000).

[21] Bernstein v. Dep't of Just., 176 F.3d 1132, 1141 (9th Cir. 1999), *reh'g granted, withdrawn*, 192 F.3d 1308 (9th Cir. 1999).

[22] *Junger*, 209 F.3d at 485.

mathematicians use equations or economists use graphs . . . . [M]athematicians and economists have adopted these modes of expression in order to facilitate the precise and rigorous expression of complex scientific ideas. Similarly, the undisputed record here makes it clear that cryptographers utilize source code in the same fashion. In light of these considerations, we conclude that encryption software, in its source code form and as employed by those in the field of cryptography, must be viewed as expressive for First Amendment purposes, and thus is entitled to the protections of the prior restraint doctrine. If the government required that mathematicians obtain a prepublication license prior to publishing material that included mathematical equations, there is no doubt that such a regime would be subject to scrutiny as a prior restraint.[23]

### A. Prior Restraint and Functionality

Importantly, the courts applied different standards to the cases. *Bernstein* applied the prior restraint doctrine, which arises when a government action prohibits speech or other expression before the speech happens.[24] Overcoming the scrutiny of a prior restraint is a tall order. The Supreme Court has noted that "[a]ny system of prior restraints of expression comes to this Court bearing a heavy presumption against its constitutional validity."[25] The government must overcome strict scrutiny by showing that (1) there is a compelling interest in the law, and (2) that the law is either narrowly tailored or is the least speech-restrictive means available to the government. In short, this means the government "carries a heavy burden of showing justification for the imposition of such a restraint."[26] There have been very few exceptions to the bar on prior restraints; only

---

[23] *Bernstein*,176 F.3d at 1141. This case was later withdrawn because the U.S. government modified its encryption regulations before the appellate court could hear the case, making the issue moot.
[24] Bantam Books v. Sullivan, 372 U.S. 58, 70 (1963).
[25] *Id.*
[26] N.Y. Times Co. v. United States, 403 U.S. 713, 714 (1971) (citing Org. for a Better Austin v. Keefe, 402 U.S. 415, 419 (1971)).

things like obscene speech, incitement to violence, and national security concerns have justified them.[27]

Instead of the prior restraint standard, the *Junger* court applied intermediate scrutiny because the law focused on the code's functionality, not its expressiveness.[28] Intermediate scrutiny consists of a two-part test: the challenged law must (1) further an important government interest (which is a lower burden than the compelling state interest required by the strict scrutiny test),[29] and (2) must do so by means that are substantially related to that interest.

Courts have applied intermediate scrutiny in other First Amendment cases and determined that for the first prong (important government interest), the government "must demonstrate that the recited harms are real, not merely conjectural, and that the regulation will in fact alleviate these harms in a direct and material way."[30] For the second prong (substantially related means), the regulation must leave "ample alternative channels of communication."[31]

More recently, in a case challenging provisions of the Digital Millennium Copyright Act (DMCA), a statute meant to prevent the circumvention of protections of access to digital content, the D.C. Circuit noted that the government "conceded that 'if you write code so somebody can read it,' it is 'expressive' speech. All of our sister circuits to have addressed the issue agree."[32] The court went on to conclude that the DMCA applied to the function of the code rather than its expression, applied intermediate scrutiny, and ruled in favor of the government.

---

[27] Near v. Minnesota ex rel. Olson, 283 U.S. 697, 716 (1931).

[28] *Junger*, 209 F.3d at 485. Had the regulation instead focused on the content of the code—its expressiveness—strict scrutiny would have likely applied which would require that the law be narrowly tailored to serve a compelling government interest. *See* Am. Libr. Ass'n v. Reno, 33 F.3d 78 (D.C. Cir. 1994).

[29] The Supreme Court has found important government interests. See, e.g., Michael M. v. Superior Ct., 450 U.S. 464 (1981) (prevention of teenage pregnancy); Craig v. Boren, 429 U.S. 190 (1976), (public health); Rostker v. Goldberg, 453 U.S. 57 (1981) (national defense); Dothard v. Rawlinson, 433 U.S. 321 (1977) (physical safety of women); Califano v. Goldfarb, 430 U.S. 199 (1977) (remediation of past societal discrimination).

[30] Turner Broad. Sys. Inc. v. FCC, 512 U.S. 622, 664 (1994).

[31] Ward v. Rock Against Racism, 491 U.S. 781, 802 (1989).

[32] Green v. U.S. Dep't of Justice, 54 F.4th 738, 745 (D.C. Cir. 2022) (citation omitted).

Given the cases discussed above, it seems clear that code is speech.[33] This means any content-neutral regulation of code, and perhaps GenAI models, would be subject to intermediate scrutiny, and an expression-based regulation would invite strict scrutiny.

### III. IMPLICATIONS

Academics, politicians, and market participants have floated various proposals to regulate large language models. This section will examine the implications of assuming the First Amendment protects GenAI models.

#### A. Regulations That May Be Prior Restraints

One often-raised proposal is to implement a licensure system whereby a regulator must certify or license a model before the developer can release or deploy it to a wide audience.[34] Other examples would be if the government prevented models from telling users how to build a bomb, make drugs, promote conspiracy theories, create malware, or access pirated movies. Such regulations may trigger the prior restraint doctrine by preventing the model from communicating. As noted above, this is a high bar, and only with the most persuasive justifications may a prior restraint pass constitutional muster.

#### B. Other Regulations That May Trigger Intermediate or Strict Scrutiny

Another common proposal is to require certain transparency thresholds for models. At the extreme, transparency requirements could include revealing the model weights and training data, but they could also include reporting impact assessments or the following kinds of information, as suggested by AI researchers at Princeton University.

---

[33] Not to stray too far from the topic of speech, but another important fact is that models are not code and are not made up of code.

[34] We acknowledge that the data collectors, data curators, model trainers, and model deployers may all be different entities. For simplicity, we treat them as a single group, and it does not affect the free speech analysis.

[F]or each category of harmful output, transparency reports must:

1. Explain how it is defined and how harmful content is detected.

2. Report how often it was encountered in the reporting period.

3. If it is the result of a Terms of Service violation, describe the enforcement mechanism and provide an analysis of its effectiveness.

4. Describe the mitigations implemented to avoid it (e.g., safety filters), and provide an analysis of their effectiveness.[35]

The issue is that courts may construe such reports as compelled speech, and compelled speech invites strict scrutiny. Most compelled speech case law arises when the government requires an entity to convey or allow a particular message or to allow space for a viewpoint the entity may disagree with.[36] That criterion does not neatly apply to models where the transparency reports do not require the model developers to carry a particular message or allow others to transmit messages through the model. Moreover, the courts take a more relaxed stance on transparency reporting when disclosure is purely factual and in a commercial context.[37]

Overall, an analysis of how the First Amendment may apply to transparency reports will require a more nuanced assessment, including examining which models the regulation would apply to (All of them? Only those of a certain size? Only those that used a certain threshold of compute? Only those that are available for certain uses or certain audiences?), whether the regulation might affect the speech rights of the model developers, and how burdensome the regulation would be to comply with. The broader the regulation, the

---

[35] Arvind Narayanan & Sayash Kapoor, *Generative AI Companies Must Publish Transparency Reports*, AI SNAKE OIL (June 26, 2023), https://www.aisnakeoil.com/p/generative-ai-companies-must-publish.
[36] *See, e.g.*, Miami Herald Pub. Co. v. Tornillo, 418 U.S. 241 (1974); Pac. Gas & Elec. Co. v. Pub. Util. Comm'n, 475 U.S. 1 (1986); Hurley v. Irish-Am. Gay Group, 515 U.S. 557 (1995); Nat'l Inst. of Fam. and Life Advoc. v. Becerra, 585 U.S. 755 (2018).
[37] *See e.g.*, Zauderer v. Off. of Disciplinary Couns. of Sup. Ct. of Ohio, 471 U.S. 626 (1985).

more likely it is to invite strict, rather than intermediate, scrutiny.[38]

### C. Takeaways

As discussed above, if models, like code, receive First Amendment protections, it could stifle or prohibit meaningful regulation of what many have claimed is a technology as powerful as fire,[39] electricity,[40] the steam engine,[41] the printing press,[42] and more.[43] Notably, none of those technologies has First Amendment protections, so none receive the same insulation from regulation. This could grant the developers of GenAI broader freedoms to experiment, innovate, and disseminate models without the fear of substantial government influence. It could also grant the developers greater influence and far more protection against government intervention than prior technologies may have enjoyed.

## IV. LET'S MENTION INTENTIONS

Many legal scholars have assumed that all GenAI outputs are speech for the sake of legal analysis.[44] If their assumptions were correct, their analyses would likely be correct, but we believe they are misplaced.

---

[38] S*ee* Netchoice, LLC v. Paxton, 49 F.4th 439 (5th Cir. 2022) (requiring disclosure of acceptable use policy likely not a violation of the First Amendment); NetChoice, LLC v. Att'y Gen., Florida, 34 F.4th 1196 (11th Cir. 2022), *vacated and remanded sub nom.* Moody v. NetChoice, LLC, 603 U.S. 707 (2024) (requiring platforms to inform users of changes to platform rules likely not a violation of the First Amendment).

[39] Prarthana Prakash, *Alphabet CEO Sundar Pichai Says That A.I. Could Be 'More Profound' Than Both Fire and Electricity—but He's Been Saying The Same Thing for Years*, FORTUNE (Apr. 17, 2023), https://fortune.com/2023/04/17/sundar-pichai-a-i-more-profound-than-fire-electricity.

[40] Shana Lynch, *Andrew Ng: Why AI Is the New Electricity*, STANFORD GRADUATE SCH. OF BUS. (Mar. 11, 2017), https://www.gsb.stanford.edu/insights/andrew-ng-why-ai-new-electricity.

[41] Hannah Levitt & Bloomberg, *JP Morgan CEO Jamie Dimon Compares AI's Potential Impact to Electricity and the Steam Engine and Says the Tech Could 'Augment Virtually Every Job'*, FORTUNE (Apr. 8, 2024), https://fortune.com/2024/04/08/jpmorgan-ceo-jamie-dimon-compares-ai-electricity-steam-engine-tech-augment-every-job/.

[42] Lauren Sforza, *Microsoft President Compares AI to Invention of Printing Press*, THE HILL (May 28, 2023), https://thehill.com/policy/technology/4024394-microsoft-president-compares-ai-to-invention-of-printing-press/.

[43] Bill Gates, *The Age of AI has Begun*, GATESNOTES (Mar. 21, 2023), https://www.gatesnotes.com/the-age-of-ai-has-begun. And these claims are not limited to for-profit organizations. Wired reports that the CEO of the nonprofit Allen Institute for Artificial Intelligence, Ali Farhadi, says he's "100 percent convinced that the hype is justified" Steven Levy, *Don't Let Mistrust of Tech Companies Blind You to the Power of AI*, WIRED (June 7, 2024), https://www.wired.com/story/dont-let-mistrust-of-tech-companies-blind-you-power-of-ai/.

[44] *See, e.g.*, Sunstein, *supra* note 6; Volokh, Lemley, & Henderson, *supra* note 6.

In cases where courts have found that code is protected speech, they have relied on several analogies: mathematical formulas, foreign languages, player piano paper, and music more generally.[45]

Upon first glance, it appears First Amendment protections could reasonably extend to model weights for a number of reasons if they can apply to mathematical equations, including that the model is essentially a compressed copy of its training data, and the training data is mostly expressive content. The model converts the expressive nature of the training data into numbers and merely transforming the format of speech does not remove its protections (e.g., making a photo a JPEG file, making a document a DOCX file, or making a song an MP3 file does not affect speech protections), or that the model is able to produce outputs that are coherent to humans.

But perhaps such an analysis gets ahead of itself by overlooking the assumption inherent in all the analogies and case law thus far: that someone intended to create the protected speech.[46] The First Amendment does not apply to random strips of black cloth (or any color, for that matter). If it did, garment factories might be in massive violation of the law every day when they discard scraps left over from sewing shirts, pants, dresses, and so on. Similarly, if the random cloth was protected, government regulations that require discarding the scraps for environmental or safety reasons might also implicate the First Amendment. As another example, if you accidentally knock over a can of paint on the sidewalk that spills onto a piece of paper, a police officer can throw the paper away and clean up the mess, ask you to clean it up, or fine you for not cleaning it up, all without implicating the First Amendment.

The reason the First Amendment can protect some pieces of cloth and some paint on paper is that they are imbued with intent by a human.[47] Intentionality, in turn, requires both sentience (the ability to feel,

---

[45] *See, e.g.*, Bernstein v. U.S. Dep't. of State, 922 F. Supp. 1426, 1435 (N.D. Cal. 1996).

[46] Consider another famous case where First Amendment protections extended to wearing a black armband to protest the Vietnam War. Tinker v. Des Moines Indep. Cmty. Sch. Dist., 393 U.S. 503, 506 (1969).

[47] Mackenzie Austin and Max Levy, *Speech Certainty: Algorithmic Speech and the Limits of the First Amendment*, 77 STAN. L. REV. 1, (2025) (making a powerful argument that the First Amendment requires "speech certainty," which is that the speaker knows what they are saying when they say it. This paper agrees with that criterion but believes requiring intentionality, including sentience and

perceive, or experience subjectivity) and self-awareness (the ability to recognize oneself as an individual).[48] Wearing a black armband was protected speech in *Tinker* because the wearer intened the cloth to convey a particular message. Likewise, if you paint a portrait on a canvas while sitting at an easel on a sidewalk, that painting is protected speech because a person intends to convey something, not merely because some paint is applied to some surface.[49]

To make the claim clearer, we can extend it to the other analogies. Music is protected because someone intended specific sounds. A player piano roll is protected because someone intended the roll to create certain sounds. Source code and object code are protected because someone intended to cause a computer to perform certain actions and communicate the intended actions to fellow coders. The same logic applies to video games and board games. In every case, a person intended for the protected speech to result in some particular message.[50] And because there is intended speech or expression, other humans have a protected right to receive the expressions.

Finally, while intention is a reasonable boundary for the sake of legal analysis--because we are willing to concede that perhaps one day machines will have true intent, and therefore their outputs could be protected-- we could also rely on a simpler human-directly-involved/no human-directly-involved dichotomy for the foreseeable future. The First Amendment has only ever recognized the speech of humans. While other creatures have appeared to make intentional communications, perhaps understanding what they are

---

self-awareness, and humanity is also necessary to ensure the definition of what could be speech is properly constrained).

[48] Even gorillas that are sentient, self-aware, and can intentionally communicate ideas that they know they are communicating when the ideas are communicated do not receive free speech protections, and people have no First Amendment right to receive gorilla communications.

[49] This does not mean the painter would win a lawsuit about being asked to move. It could be that content-neutral regulations prohibit people from painting on the sidewalk for any number of good reasons. *See, e.g.*, City of Austin v. Reagan Nat'l Advert. of Austin, LLC, 596 U.S. 61, 69 (2022) (finding that content-neutral regulations regarding billboards were not unconstitutional).

[50] Cass Sunstein mentions that the First Amendment would likely extend to Magic Eight Balls, so a government cannot say the ball must reply "Yes" when asked if a particular viewpoint is correct. Sunstein, *supra* note 6, at 11–12. How this would ever be enforced or how a company could possibly rig a Magic Eight Ball in such a way is beyond us. It seems as nonsensical as trying to restrain the speech of newborns. And this, in a way, helps make our point: we are doing too much by trying to hyperextend the First Amendment to cover more and more without a strong justification and by relying on strained analogies.

communicating when they communicate it-such as dogs whimpering for food, parrots requesting crackers, gorillas using sign language, and cats allegedly speaking English-U.S. courts, as Manheim and Atik put it "emphasize[s] the human element, the First Amendment does not protect speech as such, but only 'the freedom of speech.' Freedom is a quality that only humans enjoy. What would it mean for AI to be "free"? Free to speak? Free to believe in religion? Freedom from captivity?"[51]

GenAI lacks intentionality, sentience, self-awareness, and humanness. Therefore, unlike code and other forms of communication, nothing GenAI generates can be considered protectable speech under any reasonable reading of the Constitution or any binding case law.

## V. MODELS, UNLIKE CODE, ARE NOT SPEECH

While the lack of intention is itself dispositive of whether free speech protections should attach, we can also summarily dismiss the notion that there is a speaker by considering the only two possible speakers: the model developers and the model itself.

### A. Developers Are Not Speakers

In contrast to the examples in the preceding section regarding music and code, developers do not intend for foundation models to convey any particular message. In fact, large models–the kinds the government is most likely to regulate because they are generally more capable of producing outputs the government would be interested in regulating—are often intentionally trained not to produce a particular output because that would be less interesting and limit their usefulness to a broader audience. That is not to say the model developers have no control over the general types of outputs a model may produce, but developers do not control specific outputs.

The developers of large models also have no way of knowing how the model will associate any given tokens or how it will reply to any given input.[52] While developers may make opinionated decisions about what data to include or exclude, or what types of queries to respond to or refuse, or what kinds of outputs to filter, the model itself is merely a representation or abstraction of these choices, and it has no agency of its own. You also cannot easily "hard-code" any particular responses. As Joshua Batson, a researcher at the AI firm Anthropic, noted in a New York Times podcast,

> [t]hese models are grown more than they are programmed. So you kind of take the data, and that forms like the soil, and you construct an architecture and it's like a trellis, and you shine the light, and that's the training. And then the model sort of grows up here, and at the end, it's beautiful. It's all these little like curls, and it's holding on. But you didn't, like, tell it what to do.[53]

---

[51] Mainheim & Atik, *supra* note 9, at 169.

Instead, the outputs are merely statistically plausible tokens formed into words and sentences responding to the statistical association of the tokens created from the words and sentences of the user prompt.[54] GenAI model developers cannot know or understand when, where, or why any particular output will include any particular token.

More generally, the mere fact that a person created something does not mean that everything that entity says or does is potentially acting on behalf of the creator. Imagine if all parents were responsible for everything their teenagers said or did. Or suppose Ford was responsible for everything the eventual buyers did with the vehicles. Society has had the good sense to recognize that mere creation of something does not mean the creator always retains all rights and responsibilities associated with the thing they created.

If companies claim a restriction on output is unconstitutional, they must argue that the output is speech.

---

[52] Set aside, for the purposes of this article, the uncommon scenario of a malevolent developer who curates data to make the model more inclined to do harmful things, or who uses supervised fine-tuning or other techniques specifically to make harmful actions more likely.

[53] Kevin Roose & Casey Newton, *Google Eats Rocks, a Win for A.I. Interpretability and Safety Vibe Check*, N.Y. TIMES (May 31, 2024), https://www.nytimes.com/2024/05/31/podcasts/hardfork-google-overviews-anthropic-interpretability.html.

[54] Prompts, though they may be intricate and very creative to achieve a particular outcome, are not deterministically shaping the actual generated content. Prompt engineering optimizes the model's *likelihood* to generate what the user wants. However, whether the model actually generates what the user desires is dependent on a number of factors beyond the user's control such as the effectiveness of the model's training, the efficacy of the instruction tuning the model received, and the alignment of the model's learned human preference. Thus, the link between the prompt and generation is tenuous. Generation from an AI is intrinsically separate from user intent.

And because GenAI itself cannot create speech as it has been recognized in the U.S. legal system, the speech must be that of the company. This means all GenAI outputs would be the company's speech.

Notably, no large GenAI entity has claimed that any models represent the company's speech. If anything, they expressly disclaim this, noting that the models are experimental and that the models do not represent their views.[55] Additionally, AI developers and most users do not want to claim legal liability for anything harmful the model may output for the simple reason that they have no idea what it might output in response to any given prompt. It is also not clear whether 47 U.S.C. § 230, more commonly known as Section 230, the law that immunizes platforms from most content users post, would protect GenAI developers in the same way it shields large social media platforms, so the hesitancy is logical.[56]

### *B. Models Are Not Speakers*

The model has no intention or even understanding of what it is doing. This is largely why problems like hallucinations and shortfalls in common sense persist.[57] As discussed above, models are merely making probable guesses of which token should go next, given the prior tokens. When an AI model states a falsehood as if it is fact, it is called hallucinating. But *everything* models output is a hallucination; it just so happens that often their outputs align with reality. They have no awareness of what they are outputting, and they do not know if it is accurate or rational. GenAI cannot express itself because it has no self to express. And unlike corporations, discussed below, GenAI is not made up of humans.

A model can generate speech the way a parrot can generate speech by mimicking humans. A parrot may produce coherent outputs, but it does not have the capability to fully understand the full extent of what it is producing. A model, like a parrot, generates coherent outputs, but researchers have shown that models do not understand what they generate, no matter how sophisticated the output may appear.[58] In fact, it could be said that a parrot has even more intention than a model because it is producing output on its own initiative and can do so without prompting. Yet, nobody has claimed it would violate a person's First Amendment rights to not be allowed to listen to a parrot. Even if we assume parrots are sentient and self-aware and can speak with intentionality and speech certainty (knowing what it said when it said it), they lack the other fundamental and unavoidable characteristic of protectable speech: human origin.

---

[55] *See, e.g.*, *Gemini for Google Workspace Cheat Sheet*, SUPPORT.GOOGLE (2024), https://support.google.com/a/users/answer/14143478?hl=en (specifies that "Gemini feature suggestions don't represent Google's views, and should not be attributed to Google").

[56] *See* PETER J. BENSON & VALERIE C. BRANNON, CONG. RSCH. SERV., LSB11097, SECTION 230 IMMUNITY AND GENERATIVE ARTIFICIAL INTELLIGENCE 2–4 (2023).

[57] *See* Bender et al., *supra* note 12, at 610–23; Emily M. Bender & Alexander Koller, *Climbing Towards NLU: On Meaning, Form, and Understanding in the Age of Data*, PROC. OF THE 58TH ANN. MEETING OF THE ASS'N FOR COMP. LINGUISTICS 5185–98 (2020); Zachary Kenton et al., Alignment of Language Agents (Mar. 26, 2021) (unpublished manuscript) (https://arxiv.org/abs/2103.14659).

In fact, the same logic applies to other entities that can communicate a message that is clearly understood by humans, but that would not be protected by the First Amendment, including doorbells, thermostats, school bells, smoke detectors, and house alarms.

The reason is that the First Amendment protects speech, not the mere transmission of information. The fact that someone can ascribe meaning to something does not mean First Amendment protections automatically apply—regardless of how profound the self-imposed meaning may be. I may find the shape of a large rock in a nearby park to convey something profound about the meaning of life, but if the city decides to obscure or destroy the rock, they have not assaulted the First Amendment. I have no First Amendment right to receive the rock's expression, such that it is.[59]

The limitation on the extent of First Amendment protections is necessary. If the mere communication of information were constitutionally protected speech, then virtually nothing could be meaningfully regulated, or at the very least, it would trigger waves of incessant litigation where the government would be required to satisfy, at a minimum, intermediate scrutiny each time.

---

[58] *See* Nouha Dziri et al., *Faith and Fate: Limits of Transformers on Compositionality*, ACM DIGIT. LIBR. (2024), https://dl.acm.org/doi/10.5555/3666122.3669203; *see also* Peter West et al., *The Generative AI Paradox: What It Can Create, It May Not Understand*, ARXIV (2023), https://arxiv.org/abs/2311.00059.

[59] Note that we are not arguing that analyses and discussion about the rock or cults or religions that spring from worshiping the rock would not receive First Amendment protections. We are only concerned with the receiver/listener's supposed protections to read/view and interpret the rock, because any "insights" would derive entirely from their self-reflection. The resulting self-reflections may result in protected speech from the receiver, but no protections go from the rock to the receiver.

## C. Humans Are Not Stochastic Parrots

GenAI models are undeniably impressive as far as producing coherent outputs based on probable tokens. However, they are still only tools. They exhibit intelligence similar to the way a calculator exhibits intelligence: input by user, output by tool.[60] Yet some people still insist that human communication is merely stochastic outputs, meaning that speech is only probabilistic, and that our communication is not meaningfully different from how GenAI creates outputs. Therefore, GenAI and brains should not be treated differently under the law.[61] This is wrong.

Cognitive linguistics has long argued that speakers retain the ability to selectively compose their utterances to coincide with their communicative intent, their attitude, and the intended message.[62] Humans play an active role in organizing and personalizing what we say and how we say things. Similarly, listeners engage in an active process of construing or interpreting the received message, taking into account various contextual cues, such as shared information between the speaker and the listener as well as the intent of the speaker.

It is beyond the scope of this paper to detail all the ways human thinking, creativity, and communication are distinguishable from GenAI outputs. Still, several brief examples are worth mentioning so anyone who wishes to explore the topic further can have ideas to build on. We acknowledge that GenAI can sometimes perform some of the tasks associated with the following examples. But this reminds us of the old saying about broken clocks being correct twice a day. What follows are some of the ways in which GenAI functions is unlike how human minds work.

---

[60] And so far, calculators are better at solving math problems when told what to do with numbers. Thankfully their outputs aren't just based on the probability of, say, 2+2 equals 3 or 4 or 5.

[61] For example, Sam Altman, CEO of OpenAI, famously tweeted "I'm a stochastic parrot, and so r u[.]" Sam Altman (@sama), X (Dec. 4, 2022, 1:32 PM), https://x.com/sama/status/1599471830255177728?lang=en&mx=2.

[62] *See* Ronald W. Langacker, FOUNDATIONS OF COGNITIVE GRAMMAR: VOLUME I: THEORETICAL PREREQUISITES (Stan. Univ. Press 1987); Leonard Talmy, *Figure and Ground in Complex Sentences*, BLS (1975), https://journals.linguisticsociety.org/proceedings/index.php/BLS/article/view/2322/2092.

1. GenAI has no intentionality nor agency.[63] It does not provide any output deliberately, purposefully, or with thoughts, desires, or beliefs because it does not contain the capacity for such conditions. It merely responds to user inputs.

2. GenAI has no theory of mind.[64] It does not have any idea what you may be thinking, desiring, or believing, and it does not spend any time thinking about what you may be thinking versus what it is "thinking."

3. GenAI lacks a notion of truth or belief in what is true versus false, showing tendencies to generate false information and hallucinations.[65] It is trained to associate tokens with other tokens, not to identify truthful information from false information. It does not possess the ability to scrutinize its training data to determine whether what it was trained on is true or not, unlike how a human can question whether the information provided to us is likely true or not.

4. Language requires both form and meaning. An LLM is only trained on form (predicting the most likely next token), so it has no ability to learn or understand meaning.[66] This is why it cannot tell if something is true or false. It does not know what content is trustworthy or not.[67] Coherence does not

---

equal understanding. Past performance (a model being right about something) does not guarantee future results (that the model will continue to be correct about that topic or any other topic).

5. GenAI cannot make significant innovations.[68] In contrast, humans can create and innovate, which goes beyond mere repetition of patterns. We can compose new genres of music, invent useful technologies that have never existed before, and develop entirely new fields of study (calculus, physics, evolution, cosmology, etc.).

6. GenAI is not self-aware.[69] While humans possess self-awareness and consciousness, which allow us to reflect on our thoughts, experiences, and existence, stochastic models like GenAI entirely lack this level of meta-cognition.[70]

7. GenAI is not great at prediction and adaptation. Unlike GenAI, human learning is not just about mimicking patterns; it is about understanding principles and applying them in novel situations. We can learn from a few examples and generalize to new contexts, a trait that stochastic models struggle with because their knowledge is limited to the information they were trained on. This is why, for example, researchers found that GPT-4 did excellent on coding problems available before GPT-4's data collection cutoff date, but it performed poorly on coding problems available just after the data collection cutoff.[71] It is also why the models must be exposed to several orders of magnitude more content than humans to provide outputs that humans sometimes find useful.

---

[68] *See* Giorgio Franceschelli & Mirco Musolesi, *On the Creativity of Large Language Models* (Mar. 27, 2023) (unpublished manuscript) (https://arxiv.org/abs/2304.00008).

[69] *See* David J. Chalmers*, Could a Large Language Model Be Conscious?* (Mar. 4, 2023) (unpublished manuscript) (https://arxiv.org/abs/2303.07103) (presented at NeurIPS Conference in 2022 as an invited talk).

[70] When we asked Microsoft Copilot, powered by GPT-4, about humans being stochastic parrots it repeatedly referred to itself as being a human. So much for self-awareness.

[71] *See* Arvind Narayanan & Sayash Kapoor, *GPT-4 and Professional Benchmarks: The Wrong Answer to the Wrong Question*, AI SNAKE OIL (Mar. 20, 2023), https://www.aisnakeoil.com/p/gpt-4-and-professional-benchmarks; *see also* Ben Turner, *GPT-4 Didn't Ace The Bar Exam After All, MIT Research Suggests—It Didn't Even Break The 70th Percentile*, LIVE SCIENCE (May 31, 2024), https://www.livescience.com/technology/artificial-intelligence/gpt-4-didnt-ace-the-bar-exam-after-all-mit-research-suggests-it-barely-passed (showing OpenAI's claims about the bar exam are similarly under scrutiny).

8. Humans are deeply embedded in social and cultural contexts that implicitly shape how we understand the world. Our language and actions are influenced by these contexts in ways that are not merely stochastic. GenAI, in contrast, generally only knows how and when to adapt to a different culture if told to do so explicitly.

### D. GenAI Models Are Not Like Corporations

Some claim that GenAI should receive speech rights because other non-human entities, like corporations, receive such rights. But corporations have speech rights they are made up of humans, and all actions corporations take are on behalf of humans. This is because the language of the First Amendment is based on actions that require intention and agency. The fictional entity of "Ford" does not create advertisements; the sales and marketing teams, consisting of humans, do. Ford would not exist without humans. There is no protected corporate speech in the absence of humans.

The distinction between corporations and GenAI becomes more obvious when considering the additional rights corporations possess. For example, corporations can enter into legally binding contracts, but GenAI, like ChatGPT, cannot. Similarly, corporations can own property, but GenAI cannot.[72] It seems odd to think that GenAI cannot own something as trivial as a cup, but some people are eager to grant it powerful First Amendment rights.[73]

### E. Listeners Are Not Protected

The above sections focus on the purported speakers. But what about listeners of the alleged speech? We believe that users of models do not have a First Amendment speech right to receive model outputs. If there is no speech from the model developers and the model itself is not a speaker, then there is no speech to "listen" to or receive. If there is no speech, then there are no speech rights. However, what users of GenAI *do* with the outputs would be protected because the user would be making an intentional communication. We must separate the *generation* of outputs from the *use* of those outputs for proper legal analyses, just as the court in *Tinker v. Des Moines Independent Community School District*[74] implicitly recognized that the production of some black cloth is separate from the *use* of that cloth to protest a war. Use is protected speech.

---

[72] This may be a useful frame for thinking about when the First Amendment should apply to something. If it cannot have property rights, then it should not have speech rights.

[73] But even corporations do not have full First Amendment protections. They can be compelled to speak, for example, by SEC rules and regulations about disclosures (S-1s, 10-Ks, 8-Ks, etc.). *See e.g.*, *Form S-1*, SEC, https://www.sec.gov/files/forms-1.pdf.  The government cannot similarly compel humans to reveal all the potential risks, of, say, marrying them.

There is a related argument that people use GenAI outputs to improve their writing or to conduct research, so they should have free speech rights to what the model generates.[75] But this is misguided. Nobody is entitled to the very best of anything, including the best or easiest way to create speech. We are not entitled to a laptop with word processing software that makes composing documents and conducting research easier, and laptops do not receive First Amendment protections for merely existing or because they are more useful than a stone and chisel. GenAI is a tool, no different from pens, paper, and word processors, and using tools, by itself, does not bestow First Amendment protections on the tools themselves or give people a First Amendment right to access the tools in the condition that is most beneficial to the people. It may be that a machine gun would make a more impressive sculpture when fired on a chunk of marble than a hammer and a chisel, but we do not have an unrestricted free speech right to access and use a machine gun just because it can be used to produce what may be protected speech.

Another argument about listeners is that because people can record things and such actions are protected by the First Amendment, using a tool like GenAI also gives users First Amendment protection. But the recording rights hinge on humans intentionally creating the recording, knowing that the cameras will record what they are aimed at. Not to beat a dead horse, but there is no similar 1:1 knowledge with the input and output of GenAI. Nobody expected Google to tell people to put glue in their pizza, for example, but it

---

[74] 393 U.S. 503 (1969).
[75] *See* Volokh, Lemley, & Henderson, *supra* note 6.

happened when a user asked how to make cheese stick to the pizza.[76]

Moreover, nobody disputes that the government can regulate cameras, recorders, and other tools even though the users of those tools may have First Amendment rights from their use. All other tools that courts have granted some protection to have involved humans trying to communicate a message to other humans: the Internet, film, cable television, etc. No copper wire has been granted First Amendment protections just because it *may* be used to communicate something by transmitting signals. It is the actual *use* or attempted use that matters, not the existence of the *potential* use of a tool.

A more helpful framework would be to consider whether the information received is speech. That is, was it intentional and by a human? If it was, traditional free speech protections apply. If it was not, there are no listener rights and no first-party speech protections. Instead, it is fully non-expressive conduct.

For non-expressive conduct, the proper analysis is whether the government is attempting to forbid the recorder or listener from recording or listening, and if so, does the law improperly target and stifle some kind of downstream speech or speaker.

This is where the analysis, as applied to GenAI, becomes interesting. It cannot be the case that *any* effect on downstream speech or a speaker triggers a First Amendment analysis. If it did, anyone claiming any government action disrupted the person's speech in any manner could file a non-frivolous lawsuit. The key question becomes: at what point does something have a predictable connection to a person's potential expression?

Assuming that GenAI outputs likely impact a person's future expression, what level of judicial scrutiny should apply when the government tries to curtail GenAI outputs? We think that the rational basis test, which requires only a rational relationship to a legitimate government purpose, would be too low a hurdle for the government to meet because a law or action is generally upheld if there's any conceivable, legitimate reason

---

for it. The challenger (the person or entity challenging the law) must prove that the government has no legitimate interest or that there's no rational connection between the law and that interest, and it seems trivially easy to make up a reason to curtail some model outputs.

We also believe strict scrutiny is too high a standard. Instead, strict scrutiny should be reserved for protected speech, which GenAI cannot, by definition, produce. Therefore, the standard that makes the most sense is some form of heightened scrutiny akin to a weak version of intermediate scrutiny.

### *F. When Would Rights Attach?*

Even if one were inclined to give models some speech protections, it is unclear when those rights would attach. A model produces nonsense even after a few hundred training steps.[77] Nobody knows when, exactly, a foundation model becomes coherent. Is the gibberish output from the beginning of training protected speech?[78] If not, when do we decide to attach protections? When the output has a few understandable words? Mostly understandable words? Perfect grammar?

### *G. Summary*

When trying to identify the speaker, perhaps legal scholar Dan L. Burk described it best:

> Certainly, the machine is not a speaker for tort, First Amendment, or related purposes; as a machine, it has no awareness, cognition, or intent. Neither is the user a speaker; although the user's prompts elicit the textual output, the nature and language of the outputs are largely unanticipated and are generated by unknown (possibly unknowable) statistical mechanics. Neither is the designer, creator, or deployer of the LLM likely to be a speaker. In the case of ChatGPT, OpenAI is not aware of the details of any particular machine response, even if they may be informed of a general trend or likelihood of damaging machine responses. . . . [T]he prompter is a cause, but not a creator of the text, and the same may be said of the LLM proprietor. Consequently, LLM texts appear to entail a kind of reader response theory on steroids: essentially *all* the meaning in the text must be supplied by the reader, as there is no

---

[77] Aatish Bhatia, *Watch an A.I. Learn to Write by Reading Nothing but Jane Austen,* N.Y. TIMES (Apr. 27, 2023), https://www.nytimes.com/interactive/2023/04/26/upshot/gpt-from-scratch.html.

[78] For a helpful explainer and visualization of this process, see https://www.nytimes.com/interactive/2023/04/26/upshot/gpt-from-scratch.html.

meaning supplied by an author.[79]

In sum, there are several reasons one should conclude that GenAI outputs are typically not speech. First, the analogy often drawn from human communicative acts to describe how GenAI models work is a fundamentally misguided basis for claiming speaker-hood, as GenAI is incapable of intentions, agency, or thought. Moreover, GenAI outputs have little, if anything, in common with corporate speech, so any analogy between the two fails to accomplish much.

Even if one wanted to assign speakership to GenAI outputs, who that speaker is remains unclear. This is especially murky when the developers of GenAI are eager to disclaim the outputs of their models as their own views. Finally, it is not entirely clear when any speech rights would attach to GenAI as models display a wide variety of levels of generative capabilities depending on factors such as the model's size and training cycles. Thus, any claim of when speech rights begin would be entirely arbitrary.

For all these reasons, there is no speech. Because there is no speech, there is no need for any First Amendment speech analysis. At most, models produce non-expressive conduct, but that conduct is only protectable if a law or regulation improperly targets or stifles some probable downstream speech or speaker.

## VI. SUBSTANTIVE ARGUMENT

There may be some who read this paper and agree with the analysis that GenAI outputs are not speech under current law but still believe that courts should make the extraordinary extension of human-based free speech protections to GenAI. We believe this would be a bad idea.

### A. The Purpose of Free Speech

First, we must consider why the Constitution protects the freedom of speech. We could look at what the founders were thinking when they passed the First Amendment, but, of course, the founders could not have possibly anticipated GenAI like they could commercial speech, corporations, political speech, and the media. What we do know is that in the founding era, protected speech consisted solely of speech where the speaker was a human who spoke with intentionality, understanding what they were saying when they said it.[80] A more fruitful source for deciphering the purposes of free speech has been language provided by the Supreme Court.

---

[79] Burk, *supra* note 10, at 216–17, 222.

One of the most cited justifications for the freedom of speech is the "marketplace of ideas," which grew in part from Justice Holmes's dissent in *Abrams v. United States*.[81] He wrote that "the ultimate good desired is better reached by free trade of ideas—that the best test of truth is the power of the thought to get itself accepted in the competition of the market."[82]

Justice Brandeis, who concurred with the *Abrams* dissent, built on that idea in *Whitney v. California*[83] by championing freedom of speech as a necessary ingredient for self-governance as well. He noted that "[f]reedom to think as you will and to speak as you think are means indispensable to the discovery and spread of political truth."[84]

Nearly fifty years later, Justice Marshall made an argument geared more toward the rights associated with self-fulfillment, stating, "The First Amendment serves not only the needs of the polity, but also those of the human spirit—a spirit that demands self-expression. Such expression is an integral part of the development of ideas and a sense of identity. To suppress expression is to reject the basic human desire for recognition and affront the individual's worth and dignity."[85]

---

[80] For an even more fundamental analysis, others have ably explored the text from the founding-era definitions of "speech" and the history and the original meaning of "speech," reaching conclusions that align with this paper. See, e.g., Austin and Levy, *supra* note 47 ("…, a comprehensive survey of Founding-era dictionaries reveals remarkably consistent definitions of speech. These definitions draw sharp distinctions between thoughts and speech, defining speech as the external manifestation of something that previously existed only in the speaker's mind. That external manifestation, by its very nature, will always be capable of certain identification by its speaker...in the Founding era "there were essentially three methods of communication: oral, unamplified speech; handwritten correspondence; and printed materials created using a printing press." Each of these modes of communication are inherently and unavoidably characterized by speech certainty. By their very nature, they require that the speaker be able to identify with certainty what he said at the moment he said it. The act of speaking orally demands that something has been said aloud; writing demands something be written; printing that something be printed.")

[81] Abrams v. U.S., 250 U.S. 616, 624 (1919) (Holmes, J., dissenting).

[82] *Id.* at 630.

[83] Whitney v. California, 274 U.S. 357 (1927).

[84] *Id.* at 375.

These are not the only statements about why freedom of speech is important or perhaps even paramount to a functioning democracy, but they provide a reasonable overview of the key arguments. How, then, does GenAI fit in?

As discussed in the paper, GenAI can neither think nor offer new ideas beyond those learned in training. It is not self-aware and therefore cannot think or participate in self-expression. Thus, it is entirely unclear how granting speech protections to GenAI outputs would enhance a marketplace of ideas, how it would lead to "discovery and the spread of political truth" (GenAI has no agency and cannot investigate, interrogate, explore, or discover), or how it would lend itself to self-fulfillment (GenAI has no desires and therefore cannot wish to express itself in any particular way).

Moreover, a "marketplace of ideas" argument only works if one legal person is trying to convince another legal person that their position is correct. Humans think, form ideas, and make choices in our expressions based on various factors, including the audience and one's own stance on the topic. Therefore, we have the ability to make a case for our beliefs and convince others of them. GenAI systems, however, have none of that. They do not have an internal belief system—moral, political, or otherwise. They do not have the capacity to persuade anyone of anything, as they lack the capability of thought and self-awareness. In fact, the communications of animals like whales or house pets are closer to the speech we currently protect as they are voluntary expressions that carry meaning. Even so, we do not claim animal communications are protected speech. Even if humans are able to lend meaning to the particular meowing of their cat, that does not mean the cat's meowing becomes protected speech. If anyone is persuaded by an output by GenAI, it is because the user is affixing meaning to the output, and not because persuasion was intended by the model.

Additionally, any argument that more speech is always the cure for bad speech, and therefore we should seek to expand what speech is protected, overlooks the possibility (or the reality, really) of how GenAI can create bad speech more quickly than any human possibly could.[86] The more-speech argument came about when GenAI was unforeseeable and when humans had to craft all the speech.[87] The claim also overlooks that a tsunami of information does not enhance or facilitate any discussion—especially when it overwhelmingly springs from a single source. It is not at all clear to us how GenAI that can hallucinate a false output or be coaxed to produce outputs for the express purpose of undermining democracy at scale through various means (e.g., misinformation, disinformation, manipulation, undermining society's trust in content it encounters generally and from high-quality news sources specifically, etc.) improves our nation in a way that overwhelmingly offsets the potential and potentially irreversible harms.

---

[85] Procunier v. Martinez, 416 U.S. 396, 427 (1974) (Marshall, J., concurring).

With GenAI, there is incredible potential to be beneficial. But the benefits will not happen without human intervention. The laws of entropy teach us as much: there are more ways for things to turn out useless, to have no impact, or to be harmful than there are to be beneficial. The resting state of GenAI is not inherently beneficial. The benefits must be willed into existence and then sustained by thoughtful, concerted efforts from everyone who touches on the lifecycle, including regulators.

It is also difficult to see why a model that is incapable of safely and effectively providing medical care, hiring, or legal advice should be protected on the basis that its outputs are vital to democracy itself. It is even more difficult to understand why the First Amendment should be read to disable the government's—and therefore democracy's—ability to protect itself from non-human speech.

### B. GenAI Governance

Before handicapping the government's ability to regulate GenAI, we must also ask who else should regulate it. If not the elected government, through a democratic process with input from the elected officials or referendums of a pluralistic society, then we are choosing to cede control to a handful of unelected technologists.

---

[86] It also overlooks that GenAI models, unlike humans, cannot be dissuaded from bad speech, like defamation or "fighting words" because GenAI does not understand what it is generating.

[87] *See Whitney*, 274 U.S. at 377 ("If there be time to expose through discussion, the falsehoods and fallacies, to avert the evil by the processes of education, the remedy to be applied is more speech, not enforced silence.").

Allowing a small group of people, such as officers and directors of a handful of powerful corporations, to determine what speech is and is not acceptable for tools intended for use in nearly every profession seems anathema to the nation's pluralistic principles.[88] Unsurprisingly, the GenAI entities also lack the incentives and varied demographic and socioeconomic characteristics of the people whose democracy they are supposedly aiding. If GenAI truly is as consequential as electricity and fire, then perhaps it should be regulated as such, meaning oversight by elected officials and sometimes constrained by strict liability.

An exception to our general stance would be if U.S. citizens approve an amendment to the Constitution to grant GenAI full free speech protections. This would mean that society affirmatively chooses to expand the power of GenAI companies and constrain the power of the U.S. government to influence it. We do not think such an action would be wise, but at least it would be a democratic decision.

*C. Remaining Protections*

Suppose the argument is that if GenAI received no First Amendment protections, it would invite abuse by governments who may, for example, want to silence certain ideas. This overlooks the fact that having no speech protections is not the same as having no protections at all. It is a defining principle in the United States that viewpoint discrimination is frowned upon regardless of the First Amendment, and GenAI developers could make any number of arguments against it: due process, violation of liberty, arbitrariness, bill of attainder, and so on. It is not as if the First Amendment is the single constitutional reed protecting

---

[88] The news source with the highest subscriber count, the New York Times, has 11.4 million subscribers. https://www.nytimes.com/2025/02/05/business/media/new-york-times-q4-2024-earnings.html#:~:text=The%20New%20York%20Times%20Company%20added%20350%2C000%20digital%2Donly%20subscribers,percent%20from%20a%20year%20earlier. The GenAI system with the most subscribers, GPT-4, has 15.5 million. https://www.theinformation.com/articles/chatgpt-subscribers-nearly-tripled-to-15-5-million-in-2024. A key difference is that news sources create a limited number of articles a day. A year ago, OpenAI claimed its models were used to output 100 billion words per day. https://www.the-independent.com/tech/chatgpt-openai-words-sam-altman-b2494900.html.

everyone from an overreaching government.

As Lawrence Lessig put it when describing what he called "replicants," which are "processes that have developed a capacity to make semantic and intentional choices, the particulars of which are not plausibly ascribed to any human or team of humans in advance of those choices," he said that:

> None of this is to say that such speech is entitled to no protection at all. This is the insight in Justice Scalia's opinion in *R.A.V. v. City of St. Paul* (1992).[89] We could well conclude that replicant speech is entitled to no protection but also conclude that the government is not free to discriminate among replicant speech. From this perspective, the replicant targeting the ads in Facebook's algorithm would have no presumptive constitutional protection. But the government couldn't decide to ban Republican targeting but not targeting for Democrats. As in *R.A.V.*, that is not because the underlying speech is protected. It is because a second value within the contours of the First Amendment is the value of government neutrality.[90]

## VII. HUMANS MUST BE TREATED DIFFERENTLY FROM GENAI

While there are close calls regarding the First Amendment and when protections attach, this is not one of them. And, because models are not protected by the First Amendment, we need not consider whether regulating them may stifle someone's ability to speak or to receive speech any more than a regulation on paper or pencils or computers, all of which, like models, are mere tools and none of which, like models, are speech.

We should not go out of our way to give precious protections to entities that neither want nor need them. The fact that some listeners can ascribe meaning to model outputs is not sufficient to claim the output is speech. With models, there is no speaker and there is no intended message, so there can be no speech as it is understood in constitutional law.

Furthermore, models are not code, and, for First Amendment purposes, they are not like code. Blurring the lines for models invites a restraint on democratic governance as the government will be limited in how it can effectively control what many technology luminaries believe is one of the most consequential innovations in human history. Allowing a handful of companies to improperly rely on the powerful shield of the First Amendment when wielding such an indisputably powerful technology with far-reaching implications for democracy and the economy is unwise. A better approach is to require greater democratic participation and accountability. Whatever people may think of the developers of large language models, they probably do not think those developers have too little power.

---

[89] R.A.V. v. City of St. Paul, 505 U.S. 377, 387–96 (1992).
[90] Lawrence Lessig, *The First Amendment Does Not Protect Replicants*, in SOCIAL MEDIA, FREEDOM OF SPEECH, AND THE FUTURE OF OUR DEMOCRACY 13 (Lee C. Bollinger & Geoffrey R. Stone, eds., 2022).

Finally, if any court should feel tempted to extend free speech rights to GenAI, it must first reckon with the purpose of the First Amendment and whether protecting GenAI does more harm than good for society. Relying on strained analogies to expand legal rights is not the best way to analyze the law. However, one may define GenAI, it is clearly not a human, not an entity comprised of humans, and it is certainly not a citizen of the United States. Therefore, it makes little sense to grant it any speech rights to partake in our democratic processes.

When it comes to GenAI, which CEOs of trillion-dollar companies have compared to the most consequential technological advances in human history, the risks from too little regulation by granting virtually all GenAI outputs full free speech protections, where laws and regulations must satisfy strict scrutiny, likely dwarf the risks of too much regulation. Courts long ago recognized that having no regulations on fire or electricity would be bad for society. GenAI is no different.